# Lattice Parameters Guide Superconductivity in Iron-Arsenides

Lance M.N. Konzen,[1,2] Athena S. Sefat[2]

[1] University of California, San Diego, La Jolla, CA 92093, USA
[2] Material Science and Technology Division, Oak Ridge National Laboratory, Oak Ridge, TN 37831, USA

**ABSTRACT**

The discovery of superconducting materials has led to their use in technological marvels such as magnetic-field sensors in MRI machines, powerful research magnets, short transmission cables, and high-speed trains. Despite such applications, the uses of superconductors are not widespread because they function much below room-temperature, hence the costly cooling. Since the discovery of Cu- and Fe-based high-temperature superconductors (HTS), much intense effort has tried to explain and understand the superconducting phenomenon. While no exact explanations are given, several trends are reported in relation to the materials basis in magnetism and spin excitations. In fact, most HTS have antiferromagnetic undoped 'parent' materials that undergo a superconducting transition upon small chemical substitutions in them. As it is currently unclear which 'dopants' can favor superconductivity, this manuscript investigates crystal structure changes upon chemical substitutions, to find clues in lattice parameters for the superconducting occurrence. We review the chemical substitution effects on the crystal lattice of iron-arsenide-based crystals (2008 to present). We note that (a) HTS compounds have nearly tetragonal structures with $a$-lattice parameter close to 4 Å, and (b) superconductivity can depend strongly on the $c$-lattice parameter changes with chemical substitution. For example, a decrease in $c$-lattice parameter is required to induce 'in-plane' superconductivity. The review of lattice parameter trends in iron-arsenides presented here should guide synthesis of new materials and provoke theoretical input, giving clues for HTS.

**INTRODUCTION**

Superconductivity remains as one of the most mysterious and fascinating physical phenomenon, and there continues to be further fundamental investigations to try to understand the reasons for zero resistance and field expulsion. The two classes of high-temperature superconductors (HTS) of iron-based [1] and copper-oxide [2] materials were discovered serendipitously in 2008 and 1986, respectively, and countless efforts have synthesized varieties and analyzed many physical attributes of these materials. In fact, the microscopic understanding of HTS continues to be the central problem in condensed-matter physics, including the correlated-electron behavior of interplay between spin, charge, orbital moment, and lattice degrees of freedom. Generally, HTS arises in antiferromagnetic (AFM) so-called 'parent' materials that order below Néel ordering temperature ($T_N$). As they are chemically substituted (via dopants $x$) or pressurized [3-5], the materials' intrinsic chemical, electronic, or magnetic structures are altered, and $T_N$ can diminish to give way to a superconducting 'dome'. A hypothetical temperature-doping ($T - x$) phase diagrams demonstrates this point, demonstrated in **Fig. 1** [e.g. 6-8]. The effects of electron pairing and magnetic fluctuations seem to be crucial for causing HTS, compared to electron-phonon coupling and anomalous phonon behaviors [9]. Also, the lattice effect cannot be ignored, simply because chemical substitutions within a material can create local, hence average structural changes that greatly impact magnetism and density-of-states [10,11]. In fact, uniaxial strain [12] and anisotropic pressure [5] on parents, and isoelectronic chemical substitutions within crystals [4,13,14], may cause superconductivity. One of the remaining fundamental challenges in the field of HTS is to predict which parents, and what types and amounts of $x$ can induce superconductivity. Beyond this, the width and the height of superconducting domes (i.e. superconductivity $x$ range, and superconducting critical transition temperatures, $T_c$) should be understood.

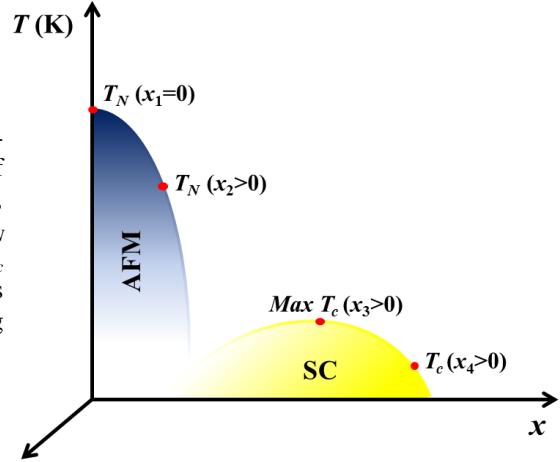

**Figure 1:** Simple representation of a generic temperature-composition ($T$-$x$) phase diagram, showing changes of correlated behavior and ordering temperature for a material, with $x$: Antiferromagnetism (AFM) and diminishing $T_N$ gives to superconductivity dome (SC) and variable $T_c$ values, depending on $x$. Such $T$-$x$ phase diagram assumes doping homogeneity within a crystalline material. Using diffraction, we study average structure changes, with $x$.

The crystal structures of all HTS are layered with two-dimensional 'in-plane' regions containing iron or copper coordination, and other 'out-of-plane' layers. A typical crystal structure is demonstrated in **Fig. 2** for $A\text{Fe}_2\text{As}_2$ parent ($A$ = alkaline-earth metal; famously called "122"). Different layers and atomic sites may be chemically substituted depending on the size, type, oxidation state, and coordination of the dopant. For example, transition metals can replace in-plane iron, oxygen can be made deficient or fluorine- or hydrogen-doped, phosphorous replaces arsenic, selenium can replace tellurium, while alkali, alkaline-earth, or rare-earth metals replace the out-of-plane filler elements (e.g. barium, yttrium, lanthanum, strontium, etc.). The doping differentiation between the different layers is important because they can influence the magnetic and superconducting phase transitions, as will be discussed below. Compared to iron-arsenides, the notion of 'in-plane' doping in cuprates is less defined, as oxygen can be inside or outside this plane, or incorporated within the numerous types of Cu-O layers. For iron-based superconductors (FeSC), there are three major antiferromagnetic parents of 122, $R\text{FeAsO}$ ($R$=rare-earth metal; "1111"), and FeTe ("11"). For cuprates, there are parents such as $\text{YBa}_2\text{Cu}_3\text{O}_6$ (YBCO) and $\text{La}_{2-x}\text{CuO}_4$ (214), although the concept of an AFM parent is also less defined for them (e.g. Tl-containing 1223, 2223 have no AFM parents). FeSC, such as 122 in **Fig. 2**, are quite versatile and 'clean' compared with the cuprates, because specific atomic sites can be targeted by synthesis in a sealed tube of controlled atmosphere [3]. In comparison, cuprates are made from reactions in air, and although historically doping to the copper site has not induced superconductivity [15-17], the levels of oxygen are hard to control and so one might suspect double-doping.

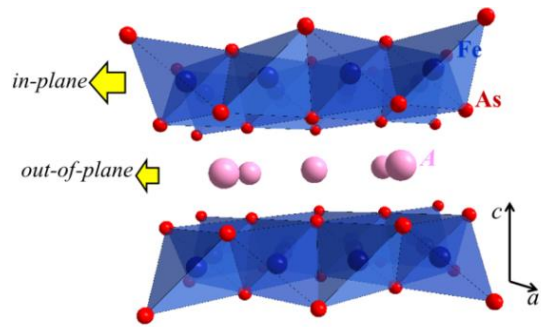

**Figure 2:** Tetragonal crystal structure of "122" with chemical formula of $A\text{Fe}_2\text{As}_2$ ($A$=Ba, Sr, or Ca). The concept of in-plane (layer containing Fe) and out-of-plane is demonstrated.

In this manuscript, we review the chemical substitution effects on the $a$- and $c$- lattice parameters in iron-based materials. In fact, a common way to find evidence of dopant substitution (i.e. stoichiometry) in a crystal is to measure the material's lattice parameter changes, using X-ray diffraction technique at room temperature. Because many HTS are manifested via certain doping ($x$) of AFM parents (**Fig. 1**), we hypothesize that the particular changes in lattice parameters with $x$ should correlate with causes of superconductivity. Hence, the results of lattice parameters with $x$ are summarized in this manuscript, which is a compilation of data of many publications.

Although superconductivity remains unpredictable, there have been two proposed structural trends in iron arsenides: (1) Lee observed that the highest $T_c$ values exist in materials with more regular $FeAs_4$ tetrahedron (close to ~109.5°) [18], while (2) Mizuguchi noted that arsenic height of $h_{As}$ ~ 1.38 Å is important for maximum $T_c$ values [19]. Fundamentally, $FeAs_4$ angle and arsenic height lend themselves to changes in average lattice parameters within a unit cell, which leads us to believe that there may be a more *simple* dependence on unit cell lattice parameter changes that can be easily detectable via quick run and refinement of X-ray diffraction data, in order to guide synthesis. Through the review of a large body of literature, we would like to add that (3) HTS can be initiated in approximately tetragonal AFM parents with $a$-lattice parameter close to 4 Å, and that (4) a decrease in $c$-lattice parameter with $x$ must be noted, if superconductivity is to occur via in-plane substitutions. We find that although there is a relationship between $a/c$ or cell volumes with $T_c$, these relationships are not consistent.

**EXPERIMENTAL AND DISCUSSIONS**

For this review paper, all of the data were extracted from tables and plots of the published manuscripts and as we reference below. In order to get the most accurate values from the literature data, the "WebPlotDigitizer" application was used. Every plot (e.g. $T$ vs $x$, lattice-parameter values vs $x$) had to be digitally cut from each publication, uploaded into the application, and each data point had to be individually identified. Over 100 plots were compiled and analyzed using this method. We must note that since many data sets were used to cross reference, we only cite original publications with clear and concise data or plots that we have used.

The data in this manuscript had to be normalized according to the initial values specified in each reported paper. Normalization was necessary because of the possibility of systematic errors in compositions and lattice parameter analyses, from publication to publication. The unit cell of the tetragonal parent material ($a_o$ or $c_o$ lattice parameters) reported in each paper was used to normalize lattice parameters for $x > 0$ materials. Without this, the plots did not overlay well, and they would be shifted due to the offsets without calibrations.

An extensive amount of time was spent exploring trends that "did not make the cut". For example, we tried to find a relationship between the lattice parameters ratio of $a/c$ with $x$ and physical behavior changes, but no consistent trends were found. In one publication, the dependence of the lattice parameters for several transition-metal doped 122-type materials were reported to depend on $a/c$ [20]. This paper gives a side by side comparison of $a/c$ for several materials. The reported change in $a/c$ is nearly identical for all dopants reviewed, however the response in the transition temperatures ($T_c$, $T_N$) for each dopant vary greatly. Although we can see that an increase in $a/c$ tends to accompany superconductivity, we have discovered a trend in $c$ alone which serves as a more concise trend that applies to more materials and properties too, as is described below. Also, taking $a/c$ obscures information; since $a$ and $c$ lattice parameters can vary independently of each other, the comparison of $a/c$ does not provide a clear representation of what is happening within the lattice. For example, when we dope a 1111-type material with fluorine [21] or hydrogen (or deuterium) [22], we see completely different responses in $a/c$, yet both transition from antiferromagnetics to superconductivity with a decrease in the $c$- lattice parameter. Fluorine doping produces a slight increase in $a/c$ while hydrogen or deuterium doping produces a slight decrease. This is because $a$-lattice is increasing in the former case, while $a$ is decreasing in the latter. Also, by looking only at $a/c$, one is unable to tell which value, $a$ or $c$, is increasing or decreasing and by what magnitude. Hence for this manuscript, we choose to analyze and focus on the simple and individual lattice parameters separately, rather than $a/c$ and volume parameter changes. We should note that trends in cell volumes can also be identified, but similar to $a/c$, volume also tends to hide details about the simple structural changes of materials as a whole.

Looking at the $a$-lattice parameters, we also did not see a trend. The $a$ remained nearly constant through many doping. In some cases, $a$ slightly went down if the $c$-lattice parameter increased, however this does not serve as a general trend to superconductivity. Likewise, any sort of behavior in $a$ did not relate to maximum $T_c$, or $T_c$ onset values. However, $T_c$ dependence with $a$-lattice parameter change can be found within individual series of compounds, for example in Hg-based cuprates [23], or in oxygen-deficient $R$FeAsO$_{1-x}$ [24].

Trends that also "did not make the cut" include using the variations (the slope) in $c$, or $a$, with $x$. In hopes of seeing a relationship between the maximum $T_c$ of a material and the rate of change in either $a$ or $c$, we took a linear fit of each parameter with $x$ and compared them. The rate of change in both $a$ and $c$ are very similar across all dopants for superconducting samples, hence we notice 'clustering' when we related it to the maximum $T_c$ values. Although the rate of change in the lattice parameters is nearly the same, their $T_c$ values differ and again, no general trend with the particular physical property was observed.

We also tried investigating a relationship within the ionic or atomic radii of the particular dopants. Since the bond lengths are affected by the size and oxidation states of the dopant, we thought that there might be some relationship to the average lattice parameters changes, and the superconducting characteristics. Unfortunately, there were no clear relations between various doping concentrations, $T_c$ onset, maximum $T_c$, size and possible oxidation states of dopants. At first, it appeared that there was a trend between electron versus hole doping in-plane. We see that $3d$ and $4d$ transition metals to the left of iron on the periodic table do not induce superconductivity while those on the right, do [20]. This would serve as a clear divide between electron and hole doping. However, no bulk superconductivity is observed in silver or gold doping of BaFe$_2$As$_2$ down to 1.8 K [25,26], and so, the effects of dopants seem to be not as simple.

For a few superconducting families, the $T_N$ and lattice parameter for the parent compounds at room temperature, and the maximum $T_c$ upon dopant $x$ are listed in **Table 1**. All HTS are tetragonal or nearly tetragonal, and have $a$ lattice parameter less than but equal to 4 Å at room temperature.

**Table 1.** A list of superconducting samples with antiferromagnetic $T_N$ at $x = 0$, maximum $T_c$ at $x$ along with lattice parameters at room temperature.

| HTS; common names | $T_N$ at $x = 0$ | maximum $T_c$, at $x$ | lattice parameters (Å) |
|---|---|---|---|
| (La$_{2-x}$Sr$_x$)CuO$_4$; 214 | 269 K | 40 K, $x = 0.15$ | $a = b = 3.786$, $c = 13.228$ |
| YBa$_2$Cu$_3$O$_{6+x}$; 123 | 420 K | 92 K, $x = 1$ | $a = 3.817$, $b = 3.885$, $c = 11.657$ |
| Ba (Fe$_{1-x}$Ni$_x$)$_2$As$_2$; 122 | 134 K | 20 K, $x = 0.05$ | $a = b = 3.964$, $c = 12.987$ |
| SmFeAsO$_{1-x}$F$_x$; 1111 | 100 K | 55 K, $x = 0.25$ | $a = b = 3.925$, $c = 8.480$ |
| FeTe$_{1-x}$Se$_x$; 11 | 70 K | 14 K, $x = 0.5$ | $a = b = 3.815$, $c = 6.069$ |

Focusing our observations on lattice parameter $c$ trends, we note that some dopants cause an increase in $c$ while others cause a decrease in $c$. An interesting observation is that when doping in-plane, only those which cause a decrease in $c$ (while maintaining a nearly constant $a$ near 4 Å) undergo a transition from antiferromagnetism to superconductivity. When *c* increases due to in-plane doping, one does not observe superconductivity. In **Fig. 3(a)** we show that doping a transition metal (T) in BaFe$_2$As$_2$ requires the *c*-lattice parameter to decrease for superconductivity to occur [7,14,20,25-32]. From further readings, this trend also holds for other in-plane doping scenarios, for example in phosphorous-doping in 122 [33] or 1111 [34], cobalt-doping 1111 [35], and selenide-doping in 11 [36] superconductors. **Fig. 3(b)** shows the response of the *c*-lattice parameter for all superconducting transition-metals doped to $A$Fe$_2$As$_2$, with $A$ = Ba, or Sr, parents [37-41]. In this figure we can see that the *c*-lattice behaves similarly for the same dopant in BaFe$_2$As$_2$ and SrFe$_2$As$_2$. For example, when doped with palladium, both parents have a steep decrease, whereas when doped with iridium, both parents see a slow decrease. Ruthenium tends to skew farther from its counterparts, perhaps because it is in the same group as iron that it is replacing.

Out-of-plane doping seems to behave unpredictably. For example, superconductivity in 1111-type occurs by replacing oxygen with fluorine [42] or hydrogen [43], and *c* decreases in both cases. Also, the rare-earth site can be replaced with decrease in *c* and cause of superconductivity [44]. However, doping alkali and alkaline-earth metals that cause superconductivity in 122 and 214-type materials experience an increase in *c* [45-49]. Doping 214-type materials with strontium increases the *c*-lattice parameter up to a certain point (about $x = 0.3$) where it then begins to decrease, much like sodium doped BaFe$_2$As$_2$ [46,49]. Hence, a decrease in *c* is necessary to induce superconductivity for in-plane doping in 122, 1111, and 11, and out-of-plane doping in 1111.

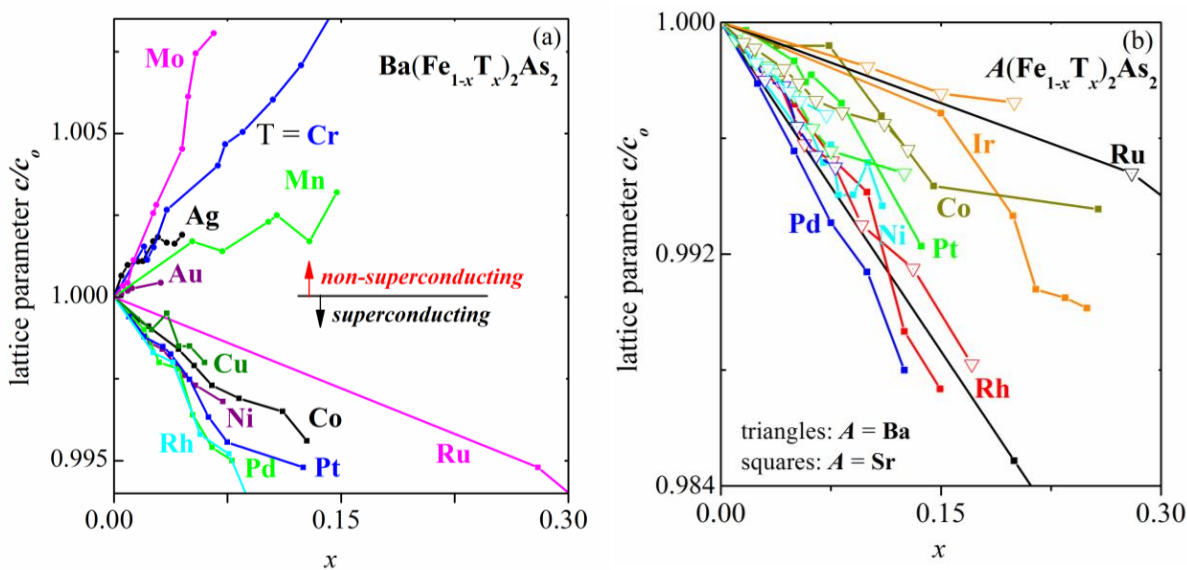

**Figure 3**: (a) Comparison of normalized *c*-lattice parameter changes for all transition-metal doped ($x$) BaFe$_2$As$_2$ [7,14,20,25-32]. (b) Comparison between *c*-lattice parameter changes for $x$ in BaFe$_2$As$_2$ (squares) versus SrFe$_2$As$_2$ (triangles) [37-41].

Since there is a strong dependence on *c,* we hypothesize that the change in *a* should also be taken into account. To incorporate both *a* and *c*, we recorded the change in volume of the unit cells as the materials are doped. **Fig. 4a** shows that there is no apparent trend in the overall volume of the crystal lattice and how it relates to SC when doping out-of-plane [45,46,50-52]. Likewise, **Fig. 4b** shows that there is no apparent trend between volume and superconductivity when doping in-plane [7,13,14,30,37,39,41,53]. In general, no trend was found relating the volume, or the rate of change in volume, to superconductivity; however, for 122-type materials we notice a relationship between the behavior of volume and the doping required to reach maximum $T_c$.

In **Fig. 5** we plot the maximum $T_c$ vs x, i.e. the peak of every superconducting dome [7,13,14,30,37,39,41,45,46,50-53]. This is the point where each material has its highest $T_c$ for any amount of *x*. We see that BaFe$_2$As$_2$ always reaches a greater maximum $T_c$ at lower doping in comparison to SrFe$_2$As$_2$. CaFe$_2$As$_2$ was not included in this trend due to a lack of publications with transition-metal doping and the variations of the parent material [54].

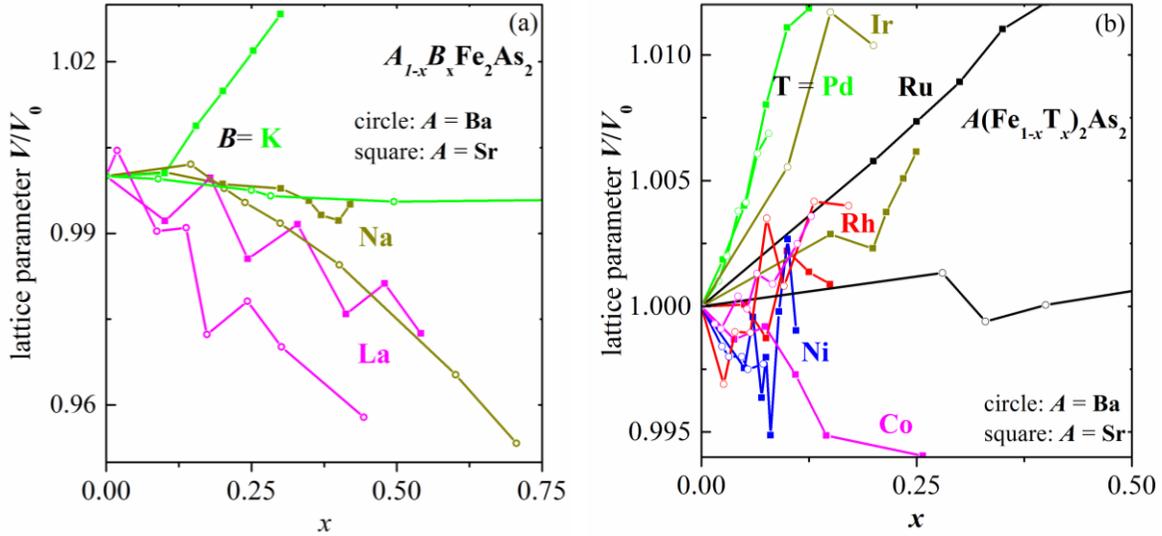

**Figure 4**: (a) Normalized volume change with doping (*x*) of 122-type materials with alkali or rare-earth metals [45,46,50-52]. (b) Normalized volume change with doping (*x*) of 122-type materials with transition metal [7,13,14,30,37,39,41,53].

When comparing the volume changes within the 122s, we see a consistent trend in relationship to doping requirements. In **Fig. 4** and **Fig. 5**, if one compares the trend in volume between different parents that are chemically substituted by the same dopant, one would note a similar spread in the doping required to reach maximum $T_c$. In other words, the volumes which deviate most from their familial counterparts also have large deviations between the doping requirements to reach their transition temperatures. We are not sure how exactly the volumes relate to max $T_c$, but there appears to be a correlation between the rate of change (slope) in *V* and the difference in doping required reaching highest $T_c$. When we note a difference between the behavior in the volume for some transition-metal doped onto SrFe$_2$As$_2$ and BaFe$_2$As$_2$, we observed that there are proportional differences in the amount of doping required to reach their respective $T_c$ max. For example, volumes that have nearly identical behavior in their rate of change when being doped to either barium 122 or strontium 122, like palladium or sodium, reach their maximum $T_c$ at exactly the same doping $x$. In this case, palladium doped to Ba-122 and Sr-122 and sodium doped to Ba-122 and Sr-122 reach maximum $T_c$ at $x = 0.1$ and $0.3$, respectively, for both parents. Volumes that have almost the same rate of change, but with some variation (nickel) reach their maximum $T_c$ at nearly the same doping $x$, but with more of a variation compared to the first. Lastly, volumes that do not have similar rates of change (cobalt, iridium, and rubidium) reach their maximum $T_c$ at very different doping $x$. Rhodium does not appear to fit this trend well and that could be due to the discontinuous behavior in its volume that may be due to chemical disorder. Hence, there is no direct relationship of the volume to maximum $T_c$, but we do see a subtle relation to doping required to reach maximum $T_c$.

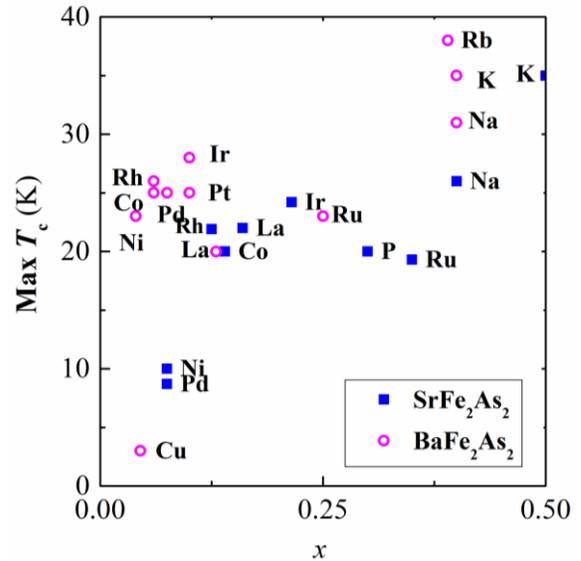

**Figure 5**: A comparison of maximum $T_c$ and doping for 122-type superconductors [7,13,14,30,37,39,41,45,46,50-53].

As stated a previous point, for in-plane substitutions, the $c$-lattice parameter must decrease for 122 to become a superconductor. In **Fig. 6** shows how the onset of superconductivity is related to the decrease in the $c$-lattice parameter in $BaFe_2As_2$ and $SrFe_2As_2$. Demonstrating the first reported x value for superconductivity (where $\rho = R = 0\Omega$) with the decrease in $c$ shows clear trends. We can conclude that the $c$-lattice parameter has to have a significant decrease, more than 0.1% of $c_o$, to initiate superconductivity for in-plane doping of 122. For $SrFe_2As_2$ (**Fig. 6b**), although the first reported $T_c$ varies greatly for each dopant, the percent decrease in $c$ falls into a fairly narrow range near 0.1% change. The range of $c$ change for $T_c$ onset for this material is noticeably larger compared to $BaFe_2As_2$. Hence, superconductivity ($R = 0\Omega$) starts with a significant decrease in $c$ which specifically for 122-type in-plane doped materials is ~0.1 - 0.4% change.

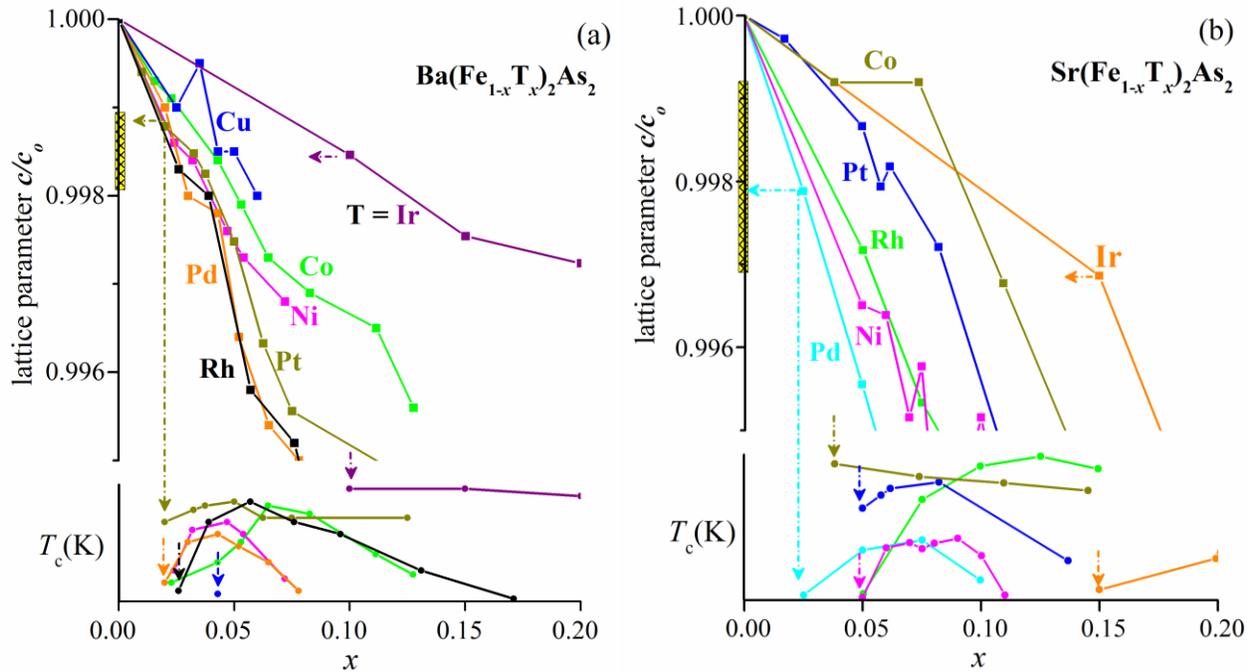

**Figure 6**: Change in $c$ with $x$, and its relation to the onset of superconductivity (shown with vertical arrows, in the lower panels of each figure). (a) Relationship between $c$ and $T_c$, with $x$, for transition-metal doped $BaFe_2As_2$ [7,20,30-32]. (b) Relationship between $c$ and $T_c$, with $x$, for transition-metal doped $SrFe_2As_2$ [37, 39-41].

**CONCLUDING REMARKS**

High-temperature superconductivity appears to be closely related to the average changes in crystal structures at room temperature. Antiferromagnetic layered materials that have an *a*-lattice parameter less than, but approximately equal to 4 Å serve as good candidates for HTS. To instigate an antiferromagnetic to superconductivity transition, a decrease in *c*-lattice with chemical substitution (or pressure) may be necessary. In fact, we have summarized that in-plane doping of 122-type iron arsenide materials requires a decrease of ~0.1 - 0.4% in *c*. For out-of-plane substitutions, the *c* trends can vary, and for example, the alkali and alkaline-earth substitutions require an increase in *c* to initiate superconductivity. This could be because we are able to dope with several types of elements (e.g. alkali, alkaline, and rare earth metals) that can hold variable sizes and oxidation states. But, such anomalies suggest that there might be some other component that better tunes and predicts superconductivity in iron-arsenide materials, or that a complicated combination of many factors (including local and average structure) matter. Our reported trends in this review manuscript should serve as a guide to further theoretical calculations and input, and may also guide and advance the chemical intuition for synthesis of additional superconducting classes and chemically-doped versions of compounds. Further work to this review manuscript may focus on other parents in addition to 122s, and also the application of specific anisotropic pressure on the crystalline lattice, for understanding of superconducting phenomena.

**ACKNOWLEDGEMENT**

This work was primarily supported by the U.S. Department of Energy (DOE), Office of Science, Basic Energy Sciences (BES), Materials Science and Engineering Division. L.M.N.K. was partially supported by U.S. DOE, Office of Science, Office of Workforce Development for Teachers and Scientists (WDTS) under the Science Undergraduate Laboratory Internship program (summer of 2016).